\documentstyle[12pt]{article}

\newcommand{\be}{\begin{equation}}
\newcommand{\ee}{\end{equation}}
\newcommand{\bea}{\begin{eqnarray}}
\newcommand{\eea}{\end{eqnarray}}

\begin{document}
\begin{titlepage}

\flushright{IP-BBSR-2013-17 }

\vspace{1in}

\begin{center}
\Large
{\bf T-duality and Scattering of Stringy States
 }

\vspace{1in}

\normalsize

\vspace{.2in}
{ Jnanadeva Maharana \footnote{Adjunct Professor, National Institute of 
Science Education and Research, Bhubaneswar, India } \\ 
E-mail maharana$@$iopb.res.in 
\\}
 
\noindent
%\today 

\normalsize
\vspace{.3in}

{\em Institute of Physics \\
Bhubaneswar - 751005 \\
India  \\ }

\end{center}

\vspace{.5in}

\baselineskip=24pt

\begin{abstract}
We present a procedure for application of $T$-duality transformation on
scattering amplitudes of closed bosonic stringy states. These states arise 
due to compactification of closed string to lower spacetime dimensions
through dimensional reduction. The amplitude, in the first quantized formalism,
is computed by introducing vertex operators. The amplitude
 is constructed by the standard
prescription and the vertex operators are required to respect conformal 
invariance. Such vertex operators are constructed in the weak field 
approximation. Therefore, the vertex operators of the stringy states of 
our interest are to be defined accordingly. We propose a prescription to 
implement $T$-duality on the three point functions and N-point functions. We
argue that it is possible to generate new amplitudes through the
transformations on a given amplitude just as T-duality transformations
can take us to a new set of string vacuum when acted upon an initial set.
Explicit examples are given for three point and four point functions.
\end{abstract}

\vspace{.7in}

\end{titlepage}

\section{Introduction }

\noindent The target space duality, T-duality, is an important symmetry
and it exhibits special attribute of string theory. We do not encounter
a symmetry of such a nature in quantum field theories. I am recapitulating
a few salient aspects of T-duality. There are excellent introductions to
T-duality in several books on string theory 
\cite{books,booksa,booksb,booksc,bookd,booke,bookf}. A number of very good
review articles \cite{rev,rev1,rev2,rev3,rev4,reva,revb,revjm}
provide pedagogical treatment of T-duality as well.

In order to
qualitatively comprehend T-duality symmetry, let us  focus our attention on
a closed bosonic string in critical spacetime dimensions, 26,
and assume that one of its spatial dimensions is compactified on a circle
of radius R. The excitations along this direction (besides the contributions
from the oscillators) will have momentum  proportional to  $1/R$ as is
the case in Kaluza-Klein compactification. However, since string is an extended
one dimensional object, it can wind around the $S^1$ and therefore,
the energy is proportional to sum of two terms (i) a term $1/R$ due to
the KK mechanism and (ii) contributions from winding term which is
proportional to $R$. The perturbative spectrum of closed string, one of whose
spatial dimension is compactified on $S^1$ with radius $R$, matches with
that of another closed string for which the corresponding spatial
coordinate is also compactified on a circle of radius $1\over R$. Thus the
spectrum of the two theories are identical and the case of latter its
KK modes and winding modes are interchanged with those of the former.
This is the $R\leftrightarrow {1/R}$ T-duality symmetry. 
In a more general compactification scheme,  
when we consider the  dimensional reduction of the 
string effective action on a torus and assume that the background 
fields do not depend
on the coordinates of the compact directions, the T-duality group 
corresponds to the noncompact $O(d,d)$ symmetry group, $d$ being the number
of compactified dimensions. Thus given a set of backgrounds satisfying the
constraints imposed by worldsheet conformal invariance, one can judiciously
implement $O(d,d)$ transformation and generate a new set of inequivalent
 backgrounds. These backgrounds are such that they cannot be connected to
the original backgrounds through 'gauge transformations' i.e. through
general coordinate transformations associated with the target space metric
$G$ or the gauge transformation of the two form potential $B$, if the
effective action is obtained from the worldsheet $\sigma$ model in the 
presence of the backgrounds $G$ and $B$. The effective action is derived
through the computation of the $\beta$-function of the $\sigma$ model where
the backgrounds  $G$ and $B$ play the role of coupling constants. 
Moreover, as is well known,
 T-duality has played a very important role in study of string cosmology,
in black hole physics and in the context of $D_p$-branes. 
\\
The target space duality has been investigated from the perspective of the
string worldsheet. I advocate the point of view that this approach provides
us a deeper understanding of the symmetry since this duality is intimately
related to the fact that string is a one-dimensional object. As mentioned
above when we toroidally compactify some spatial dimensions and dimensionally
reduce the effective action, the  resulting lower dimensional action is 
expressed in manifestly $O(d,d)$ invariant form. When we resort to the 
worldsheet description, some careful analysis is necessary in order
to reveal the underlying $O(d,d)$ symmetry. The standard Polyakov
worldsheet action of the 
compactified theory is not invariant under $O(d,d)$ transformations. 
 Indeed, it is customary to  introduce
dual (string) coordinates for the compact coordinates (which satisfy the same
worldsheet boundary conditions as the original compact coordinates)
and dual backgrounds \cite{duff,mahat,ms}
(these are independent of compact coordinates)
which live along compact directions. Thus we construct a dual action in terms
of the dual compact coordinates and the corresponding 
dual backgrounds. The 'spacetime'
string coordinates and the backgrounds (which transform as spacetime tensors)
are inert under T-duality thus this part of the worldsheet action is retained
as it is. Next, one obtains the Euler-Lagrange's equations of motion for
the compact string coordinates in the presence of backgrounds as well as the
equations of motion resulting from dual action. If the number of compact
coordinates is $d$, then we have, in total,  $2d$ set of equations of motion.
The two sets of equations of motion can be suitably rearranged to cast them
 in an $O(d,d)$ covariant form. The results derived for closed
bosonic string can also be derived 
 for NSR superstring with certain modifications
and technical improvements \cite{jmnsr} over previous works
\cite{dm,warren,rest,resta,restb,restc}.\\
It is evident that T-duality is an important symmetry and it has played a very
important role to generate new background configuration from known ones.
Especially, it has been utilized very crucially in study of string
cosmology \cite{reva,revb} and in the physics of stringy back holes
from diverse directions \cite{duff1,youm,just}.\\
The purpose of this article is to exhibit how T-duality can be applied to
study scattering of string states. 
Let us very briefly summarize the steps for constructing scattering 
amplitudes for the closed string states in the first quantized formalism. The
first ingredient is the vertex operator associated with the stringy states. It
is required to satisfy a set of general principles (see Section 3). These 
vertex operators are generally constructed in the weak field approximation;
for example, graviton is a massless excitation. Thus for the graviton, in this
approximation, we consider a weak field around the Minkowski metric.
Therefore,  the vacuum is the free string vacuum and all correlators/Green's 
functions are evaluated in this frame work. As a consequence, in computation
of three point function or any N-point functions, involving any of the stringy
states, we have the product of the corresponding set of normal ordered vertex
operators. Then one can adopt any of the well defined formalisms to compute
the scattering amplitudes.
As alluded to above,
we can generate new backgrounds from known backgrounds by implementing
$O(d,d)$ transformations. In this proposal, we intend to generate new
scattering amplitudes from a given amplitude with judicious implementation
$T$-duality symmetry. Therefore,in order to achieve this goal, 
we have to adopt the weak field
approximation. \\
We outline the prescription below. Let us consider the moduli $G$ and $B$ 
arising from the dimensional reduction of the graviton and the antisymmetric
tensor field from the two massless states of a closed string in (say)
critical dimensions. The duality transformation rules for the each of the
individual scalars  $G$ and $B$  separately is quite involved; even that
of $G+B$ is not very simple either. Therefore, in order to study the
emergence of $T$-duality symmetry, $O(d,d)$ when it is compactified on $T^d$,
it is very convenient to introduce the so called $M$-matrix (see the next 
Section for more details). The new backgrounds, say $G'$ and $B'$, are
generated by first implementing $T$-duality transformation on $M$-matrix
(to get an $M'$-matrix) and extract  $G'$ and $B'$ from  $M'$-matrix.\\
In the context of scattering of moduli and/or their counter parts arising
from dimensional reduction of excited massive level, first we have to lay down
the procedure.  We adopt the weak field approximation for the backgrounds
(say  $G$ and $B$). Next task is to construct the corresponding vertex
operators in this frame work. However, if we intend to study duality 
transformations on the desired scattering amplitude, the vertex operators
as such associated with  $G$ and $B$, are not quite suitable. Therefore,
I feel that we must first construct the corresponding $M$-matrix and 
incorporate the weak field approximation to this matrix to begin with.
I have introduced a proposal to construct the desired vertex operators
originally starting from the $M$-matrix itself. I argue that, within my
prescription, we can implement $T$-duality on three point and higher point
functions since it acts on the vertex operators. Thus starting from a given
amplitude a new amplitude can be generated by implementing $T$-duality
operations.\\   
The rest of the article is 
organized as follows. In Section 2, we examine how $O(d,d)$ transformations
are implemented for weak background fields and study their consequences.
Section 3 contains some applications of our proposal. First, we consider
a 3-point  function, consisting of three vertex operators of massless states.
These states arise from compactification of $\hat D$-dimensional graviton
to lower dimension and these 
external legs are on shell. Then we argue how this vertex can be transformed
under T-duality describing another 3-point function. 
I have studied T-duality
symmetry associated with excited massive states of closed bosonic string
\cite{jmmass,revjm}.
I construct the three point vertex involving the first 
 excited massive state vertex
operator and two other vertex operators for scalars arising from 
dimensional reduction of the graviton. I also give an explicit example
of $T$-duality transformation when we compactify on $T^2$ and the
corresponding duality group is $O(2,2)$. 
There is a formalism due to Kawai, Llewellen and
Tye \cite{klt} where they demonstrate that the tree level closed string
N-point amplitudes can be expressed in a factorized form in terms of open 
string N-point amplitudes. I utilize this formalism and propose a new way
to implement T-duality transformations on closed string amplitude when the 
string is compactified to lower $D$ spacetime dimension with $d$ compactified
dimensions. In particular,this formalism looks quite promising if we consider
higher excited massive states and their tree level amplitudes
 Then I consider
a 4-point amplitude to implement T-duality in weak fiend approximation for
tachyon and a gauge boson. This gauge boson arises due to the isometries
associated with dimensional reduction. The $T$-duality transformation emerges 
elegantly when we appeal to \cite{klt} with appropriate modifications suitable
for our proposal in this work.
I  present an argument how my proposal will work for scattering involving 
massive excited state with other states. 
Section 4 contains discussions and conclusions.

\section{T-duality in Weak Background Approximation}

\bigskip
\noindent Let us first recapitulate some of the salient features of T-duality
from the worldsheet point of view. Consider evolution of a closed bosonic 
string in $\hat D$ spacetime dimensions in the background of its massless
excitations, graviton and antisymmetric fields denoted by $\hat G$ and
$\hat B$.
\bea
\label{poly}
S= {1\over 2}\int d^2\sigma \bigg(\gamma^{ab} 
\partial _a{\hat X}^{\hat\mu}\partial _b{\hat X}^{\hat\nu}
{\hat G}_{\hat\mu\hat\nu}({\hat X}^{\hat\mu})+
\epsilon^{ab}\partial _a{\hat X}^{\hat\mu}\partial _b{\hat X}^{\hat\nu}
{\hat B}_{\hat\mu\hat\nu}({\hat X}^{\hat\mu})\bigg)
 \eea
where $\gamma^{ab}$ is the worldsheet flat (inverse) metric. $\hat D=0,1..
{\hat D}-1$. The backgrounds depend on string coordinates ${\hat X}^{\hat\mu}$.
This is identified as the two dimensional $\sigma$ model action. 
${\hat G}_{\hat\mu\hat\nu}({\hat X}^{\hat\mu})$
and ${\hat B}_{\hat\mu\hat\nu}({\hat X}^{\hat\mu})$ are to be treated as
coupling constants.
In order to get an intuitive perspective, let us assume,
for the time being that  $\hat G$ and $\hat B$
are independent of ${\hat X}^{\hat\mu}$. The canonical Hamiltonian density is
expressed in terms of ${\hat M}$-matrix (see below)
\cite{mmatrix,matrixa,matrixb,matrixc,matrix2,matrix2x,matrix2a,matrix2ay}
\bea
H_c=\frac{1}{2}\pmatrix{{\hat P} & {\hat X}' \cr}
\pmatrix{ {\hat G}^{-1} & -{\hat G}^{-1}{\hat B} \cr
             {\hat B}{\hat G}^{-1} & {\hat G}-{\hat B}{\hat G}^{-1}{\hat B} \cr}
\pmatrix{ {\hat P} \cr {\hat X}' \cr}
\eea
where $\hat P$ are conjugate momenta of ${\hat X}^{\hat\mu}$.  The Hamiltonian 
density, $H_c$ is invariant under global $O({\hat D}, {\hat D})$ 
transformations
\bea
\pmatrix{ {\hat P} \cr {\hat X}' \cr} \rightarrow 
\Omega_0 \pmatrix{ {\hat P} \cr {\hat X}' \cr} 
\eea
$\hat M$ is symmetric and transforms as 
\bea
\label{transodd}
 {\hat M}\rightarrow\Omega_0{\hat M}\Omega^T_0 , ~~
\Omega^T_0\eta_0 \Omega_0=\eta_0,~~\Omega_0\in O(\hat D, \hat D)
\eea
where 
\bea
{\hat M}= \pmatrix{ {\hat G}^{-1} & -{\hat G}^{-1}{\hat B} \cr
         {\hat B}{\hat G}^{-1} & {\hat G}-{\hat B}{\hat G}^{-1}{\hat B} \cr}
\eea
and $\eta_0$ is the  $O(\hat D, \hat D)$ metric.
\be
\eta_0=\pmatrix{ 0 &  {\bf 1} \cr {\bf 1}  & 0 \cr}
\ee
where  $\bf 1$ is  $ {\hat D}\times {\hat D} $ unit matrix and 
${\hat M}\in O({\hat D},{\hat D})$.
Let us consider the scenario when the ${\hat{\cal M}}_{\hat D}$-dimensional
manifold is decomposed such that we have a lower dimensional spacetime
manifold ${\cal M}_D$ and a compact manifold $K_d$ .i.e.
${\hat {\cal M}}_{\hat D}={\cal M}_D\otimes K_d$. Let the coordinates on
${\cal M}$ be denoted by $X^{\mu}, \mu=0,1..D-1$ and those on $K_d$ by
$Y^{\alpha}, \alpha=1,2,...d$ so that ${\hat D}=D+d$.
Further more, in adopting Scherk-Schwarz \cite{ss} dimensional reduction, 
it is assumed, in this work, 
that all the backgrounds depend on spacetime coordinates $X^{\mu}$ only.
When we examine the worldsheet action (\ref{poly}), we note that all
backgrounds, when appropriately decomposed, depend only on the string
coordinate $X^{\mu}$. Next we decompose the ${\hat D}$-dimensional tensors
$\hat G$ and $\hat B$ to lower dimensions \cite{ms}
and as is well known,  we get
tensors, vectors and scalars in lower spacetime dimension $D$. The vierbein
formalism for the metric is convenient for our purpose and the vierbein 
which defines the $\hat D$-dimensional metric is given by 
\bea
\label{schwarz}
 {\hat e}^{\hat r}_{\hat\mu}=\pmatrix{ e^r_{\mu}(X) &
A^{(1)\beta}_{\mu}(X)E^a_{\beta}(X)\cr 0 & E^a_{\alpha}(X)\cr }
\eea
The spacetime metric is $g_{\mu\nu}=e^r_{\mu}g^{(0)}_{rs}e^s_{\nu}$ and
the internal metric is $G_{\alpha\beta}=E^a_{\alpha}\delta_{ab}E^b_{\beta}$;
$g^{(0)}_{rs}$ is the D-dimensional flat space Lorentzian signature metric.
$ A^{(1)\beta}_{\mu}$ are gauge fields associated with the d-isometries and
it is assumes that the backgrounds depend on coordinates $X^{\mu}$ and are
independent of $Y^{\alpha}$. Similarly, the antisymmetric tensor background,
depending only on $X^{\mu}$ can be decomposed as
\bea
 {\hat B}_{{\hat\mu}{\hat\nu}} =\pmatrix{B_{\mu\nu}(X) & B_{\mu\alpha}(X)\cr
B_{\nu\beta}(X) & B_{\alpha\beta}(X) \cr}
\eea
As expected, there are gauge fields, $ B_{\mu\alpha}(X)$, arising from 
the compactification of $\hat B$. After the dimensional reduction, we
also encounter the $M$-matrix which  belongs to T-duality group $O(d,d)$.
It is given by 
\bea
\label{matrixm}
M=\pmatrix{ G^{-1} & -G^{-1}B \cr
  BG^{-1} & G-BG^{-1}B \cr}
\eea
Note that $M \in O(d,d)$ and therefore, it satisfies the usual constraints,
$M\eta M =\eta$ and $\eta$, the  $O(d,d)$ metric is
\be
\eta = \pmatrix{0 & 1\cr 1 & 0 \cr}
\ee
and $1$ is $d \times d$ unit matrix. Under the $O(d,d)$ transformation,
$M\rightarrow \Omega M\Omega^T$ and the metric $\eta$ remains invariant.
In order to generate new nontrivial backgrounds, starting from an initial
set of backgrounds, we apply desired $O(d,d)$ transformations.
For sake of simplicity, let us set $A_{\mu}^{\alpha}=0$ and  
$ B_{\mu\alpha}(X)=0 $ (note that these two gauge field can be combined
following the rules of dimensional reduction and this combination of the 
gauge fields transforms as a vector under 
$O(d,d)$ transformations \cite{ms}). 
The the Hamiltonian density corresponding to internal
coordinates $\{ Y^{\alpha} \}$ is given by 
\bea
\label{hamiltonian}
H_c^d=\frac{1}{2}\pmatrix{ P & Y' \cr}M\pmatrix{ P \cr Y' \cr}
\eea
$P_{\alpha}= G_{\alpha\beta}{\dot Y}^{\beta}+
B_{\alpha\beta}Y'^{\beta}$ are the momenta conjugate to $Y^{\alpha}$,
here 'dot' and 'prime' denotes derivative with respect to the 
worldsheet coordinates $\tau$ and 
$\sigma$ respectively.
The  vertex operator associated with $G_{\alpha\beta}(X)$ can be identifies as 
$G_{\alpha\beta}(X)\partial Y^{\alpha}{\bar \partial}Y^{\beta}$. 
In terms of phase space variables it is:  
$G^{\alpha\beta}(X)P_{\alpha}P_{\beta}+G_{\alpha\beta}(X)Y'^{\alpha}Y'^{\beta}$.
If we intend to compute, for example, scattering of two of these massless
particles, we have to resort to approximation scheme as has been discussed
in the previous section. 
The conventional method is to
adopt weak field approximation. As an illustrative example, consider vertex
operators for spacetime graviton. Here we write
\be
g_{\mu\nu}=g^{(0)}_{\mu\nu}+a{\tilde h}_{\mu\nu}(X)
\ee
where $g^{(0)}_{\mu\nu}$ is the flat Minkowski space metric and 'a'
is a small expansion parameter (is related to Newton's constant in linearized
approximation of Einstein-Hilbert action) and ${\tilde h}_{\mu\nu}(X)$ is the
graviton in the weak field approximation . Then the graviton vertex is
\be
\label{gvertex}
{\tilde h}_{\mu\nu}(X)\partial X^{\mu}{\bar\partial}X^{\nu}
\ee
and the canonical momentum is defined from the free string Lagrangian,
$P_{\mu}={\dot X}^{\mu}$; moreover, $\partial X^{\mu}={\dot X}^{\mu}+X'^{\mu}$
and ${\bar\partial} X^{\mu}={\dot X}^{\mu}-X'^{\mu}$. 
The vertex operator (\ref{gvertex}) is required
to be $(1,0)$ and $(0,1)$ primary 
with respect to the free string stress energy momentum
tensors, $T_{++}$ and $T_{--}$ respectively and these are  given by
\be
\label{xtt}
T_{++}={1\over 2}\partial X^{\mu}\partial X^{\nu}g^{(0)}_{\mu\nu},~~~~
T_{--}= {1\over 2}{\bar\partial}X^{\mu}{\bar\partial}X^{\nu}g^{(0)}_{\mu\nu} 
\ee
Now we quote the standard result when we demand (\ref{gvertex}) to have
conformal weight $(1,1)$.
\be
\nabla^2 {\tilde h}_{\mu\nu}(X)=0, ~~~~
 \partial^{\mu}{\tilde h}_{\mu\nu}(X)=\partial^{\nu}{\tilde h}_{\mu\nu}(X)=0
\ee
Here $\nabla^2$ is the flat space D-dimensional Laplacian.
If we express ${\tilde h} _{\mu\nu}=e^{ik.X}\epsilon_{\mu\nu},~ 
\epsilon_{\mu\nu}$ being
the polarization tensor of graviton. The above constraints translate to
the familiar equations,
\be
k^2=0,~~~~ k^{\mu}\epsilon_{\mu\nu}=  k^{\nu}\epsilon_{\mu\nu} =0
\ee
Similarly, we can write
\be
B_{\mu\nu}= b^{(0)}_{\mu\nu}+a {\tilde b}_{\mu\nu}(X)
\ee
and $b^{(0)}_{\mu\nu}$ is the constant antisymmetric tensor, analog of 
$g^{(0)}_{\mu\nu}$. However, presence of this constant antisymmetric 
tensor in the free string
worldsheet action, does not contribute to the equations of motion (we are
dealing with noncompact string coordinates) and we can still decompose
$X^{\mu}(\sigma , \tau)= X^{\mu}_L+ X^{\mu}_R$. Thus we can ignore the 
presence of  this constant tensor  in all our subsequent discussion. Moreover, 
${\tilde b}_{\mu\nu}(X)$ also satisfies similar
constraints as ${\tilde h}_{\mu\nu}(X)$ as above
\be
\nabla^2 {\tilde b}_{\mu\nu}=0,~~~~~ 
\partial^{\mu}{\tilde b}_{\mu\nu}=\partial^{\nu}{\tilde b}_{\mu\nu}=0
\ee
Similarly, for the backgrounds $G_{\alpha\beta}$ and $B_{\alpha\beta}$ we adopt
the weak field approximation
\be
\label{weakgb}
G_{\alpha\beta}=\delta_{\alpha\beta}+ah_{\alpha\beta},~~{\rm and} ~~~
B_{\alpha\beta}=ab_{\alpha\beta}
\ee
and '$a$' being the expansion parameter. Although we might loosely use the
term toroidal compactification, it is understood that we do not account
for special features of winding modes etc in this investigation.
Note that $G_{\alpha\beta}$ and $B_{\alpha\beta}$ transform as scalars (these
are moduli) from the point of view of $D$-dimensional spacetime and these
are symmetric and antisymmetric under $\alpha\leftrightarrow\beta$ respectively.
Moreover, $h_{\alpha\beta}$ and $b_{\alpha\beta}$ also share the same 
attributes. If we demand the vertex operators 
$h_{\alpha\beta}\partial Y^{\alpha} {\bar \partial} Y^{\beta}$ and
$b_{\alpha\beta}\partial Y^{\alpha} {\bar \partial} Y^{\beta}$ to be $(1,1)$
primaries with respect to 
\bea
T_{++}={1\over 2}\partial Y^{\alpha}\partial Y^{\beta}\delta_{\alpha\beta},~~
T_{--}= {1\over 2}{\bar\partial}Y^{\alpha}{\bar\partial}Y^{\beta}
\delta_{\alpha\beta}
\eea
then we arrive at
\be
\nabla^2 h_{\alpha\beta}=0,~~{\rm and }~~ \nabla^2 b_{\alpha\beta}=0
\ee
We remark in passing that if we had considered the gauge fields
$A^{\alpha}_{\mu}$ and $B_{\alpha\mu}$ they will couple to the worldsheet
coordinates $X^{\mu}$ and $Y^{\alpha}$. The conformal invariance implies
equations of motion
\be
\nabla^2 A^{\alpha}_{\mu}=0,~~ {\rm and} ~~~\nabla^2 B_{\alpha\mu}=0
\ee
and the transversality conditions
\be
\partial^{\mu}  A^{\alpha}_{\mu}=0,~~~ {\rm and} ~~~
\partial^{\mu} B_{\alpha\mu}=0
\ee
If we had envisaged the ${\hat D}$-dimensional worldsheet action with 
backgrounds $\hat G$ and $\hat B$, the conformal invariance will lead to
equations of motion of the form
\be
\label{oldgvertex}
{\hat \nabla}^2{\hat h}=0, ~~  {\rm and} ~~~{\hat\nabla}^2{\hat b}=0
\ee
where ${\hat h}$ and ${\hat b}$ are to be understood as weak field expansions
of $\hat G$ and $\hat B$. We also obtain
the  appropriate transversality conditions on these background. Note that,
upon dimensional reduction ${\hat G}\rightarrow \{g_{\mu\nu}, A^{\alpha}_{\mu},
G_{\alpha\beta}\}$ and ${\hat B} \rightarrow \{B_{\mu\nu}, B_{\alpha\mu},
B_{\alpha\beta} \}$ and the conditions we have derived above are the same i.e.
the tensors and gauge fields satisfy equations of motion and transversality
conditions. Whereas the moduli satisfy equations of motion as given above.
In the Hamiltonian formulation, the vertex operators can be obtained too.
Moreover, in the weak field approximation the canonical momenta are
taken to be the $\tau$ derivative of the corresponding string coordinates.
If we focus attention on the compact coordinates $\{ Y^{\alpha} \}$ and 
assume they couple to corresponding backgrounds (we confine to massless
excitations) which only carry $X^{\mu}$ dependence then we could explore
T-duality properties of the vertex operators. Since the Hamiltonian density
(\ref{hamiltonian}) is T-duality invariant, we may obtain nontrivial new
backgrounds by implementing T-duality transformations and the string
will propagate in these backgrounds. These backgrounds are desired to fulfill
constraints of conformal invariance. However, if we intend to consider
interactions such as three point and four point functions (or even 
N-point functions), then the vertex
operators obtained from (\ref{hamiltonian}) are not useful and we should
resort to the weak field approximation.
\\
In view of preceding remarks, I propose to define the $M$-matrix 
(\ref{matrixm}), in the weak field approximation, as follows
\be
\label{wfield}
M={\bf 1}+a{\tilde M}
\ee
where ${\bf 1}$ is the $2d \times 2d$ unit matrix, '$a$' is the small 
expansion parameter. Sometimes we omit the presence of of '$a$' in such an
expansion. However, it is always understood that such  weak field 
'fluctuations' are due to the presence of $'a'$ even if it does not appear
explicitly. 
The backgrounds $G^{-1}$, $G$ and $B$ appearing in
the definition of $M$, (\ref{matrixm}) have weak field expansions
(\ref{weakgb}) as alluded to earlier. Therefore, $\tilde M$ can be
expressed in terms of $ h^{\alpha\beta}$,
$h_{\alpha\beta}$ and $b_{\alpha\beta}$ (note that  $ h^{\alpha\beta}$ and
$h_{\alpha\beta}$ are related and see below). 
Since $M \in O(d,d)$ it satisfies the constraint $M\eta M=\eta$, 
\be
\label{mtilde}
{\tilde M}\eta + \eta{\tilde M}=0
\ee
Note that $\tilde M$ is not an $O(d,d)$ matrix unlike $M$. Furthermore,
from the relation $G_{\alpha\beta}G^{\beta\gamma}=\delta^{\gamma}_{\alpha}$,
we get
\be
h^{\alpha\beta}=-\delta^{\alpha\gamma}h_{\gamma\sigma}\delta^{\sigma\beta}
\ee
when we retain terms to order '$a$' in expansion (\ref{wfield}). The above
relation between $h^{\alpha\beta}$ and $h_{\alpha\beta}$ is consistent with
the constraint (\ref{mtilde}).
Moreover, by retaining terms to order '$a$', we notice that block the diagonal
elements of $\tilde M$ are $h^{\alpha\beta}$ and 
$h_{\alpha\beta}$ and the two off diagonal
block elements are $-b$ and $b$. Thus $\tilde M$ is also a symmetric matrix like
$M$-matrix; however, it has to satisfy (\ref{mtilde}).\\
Now I present the steps as how to generate new set of backgrounds from  given
backgrounds. Let us start with initial backgrounds $G_{in}$ and $B_{in}$ and
define the matrix $M_{in}$ and corresponding ${\tilde M}_{in}$ in order to 
implement the weak field approximation. Let $\Omega $ be the desired $O(d,d)$
transformation. Then the new matrix, $M_{f}$ is given by
\be
\label{mf}
M_{in}\rightarrow M_{f}=\Omega M_{in}\Omega^T
\ee
I adopt the prescription that just as $M_{in}$ has weak field approximation
$M_{in}={\bf 1}+{\tilde M}_{in}$, having obtained $M_{f}$, it will have weak
field expansion $M_{f}={\bf 1}+{\tilde M}_{f}$. In order to decompose
$M_{f}$ in this manner we might have to rescale the new metric appearing in the
definition of $M_f$  obtained through
(\ref{mf}). Let us explore more about $\tilde M$. Recall that \cite{ms} we
can write an $O(d,d)$ matrix
\be
\Omega=\pmatrix{ \Omega_{11} & \Omega_{12} \cr \Omega_{21} & \Omega_{22}\cr}
\ee
satisfying $\Omega^T\eta \Omega=\eta$. With this decomposition $G+B$ transform
as quotients reminiscent of $SL(2,R)$ transformation and this avenue does
not take us too far to get handle on new backgrounds as can be seen by working
out the details and try to carry out the weak field approximation for the
transformed $G+B$. Therefore, we incorporate the T-duality transformation
through the $\Omega$ transformation on the $M$-matrix.
Let us assume that under T-duality
\be
\label{oddtrans}
\pmatrix{ G^{-1} & -G^{-1}B \cr
  BG^{-1} & G-BG^{-1}B \cr}
\rightarrow 
\pmatrix{ G'^{-1} & -G'^{-1}B' \cr
  B'G'^{-1} & G'-B'G'^{-1}B' \cr}
\ee
We can obtain $G'$ by inverting top left block matrix $G'^{-1}$ and then
easily extract antisymmetric $B'$ matrix. As mentioned earlier, then
carry out the weak field approximation. Notice that $G'$ and $B'$ will be
expressed in terms of $G$ and $B$ once we specify an $O(d,d)$ transformation
(see the next section for an explicit example). It is obvious from examination
of (\ref{oddtrans}) that the right hand side off diagonal block matrix will
eventually lead to extraction of nontrivial $B'$ in general.
Now we are in a position
to implement T-duality on scattering amplitudes.

\bigskip

\section{Scatterings in the Weak Field Approximation}

\bigskip

\noindent In this section we explore how one can apply T-duality transformation
on  a scattering 
 amplitude to relate it to another one. Our strategy is as follows. Let us
consider a three point function i.e. product of three vertex operators.
We start with a vertex operator, call it $V^{(1)}$, 
with only $h_{\alpha\beta}\partial Y^{\alpha}{\bar\partial} Y^{\beta}$ and two
other vertex operators. Next, we decide to 
apply T-duality on $V^{(1)}$, then under 
this transformation $h_{\alpha\beta}\rightarrow h'_{\alpha\beta}$ and it 
can have an additional  term  
$b'_{\alpha\beta} \partial Y^{\alpha}{\bar\partial} Y^{\beta}$
although the antisymmetric field  $b_{\alpha\beta}$ was absent to start
with. This is achieved  follows  according to our proposal . 
Given initially an 
 $h_{\alpha\beta}$ we construct the corresponding ${\tilde M}_{in}$. 
Then we implement
the $O(d,d)$ transformation on $M_{in}$ constructed from ${\tilde M}_{in}$
to generate $M_{f}$. Thus  ${\tilde M}_{f}$ can be defined once $M_f$
is generated and finally we shall extract 
 $b'_{\alpha\beta}$. This procedure is an application for compactification on
$T^d$ (we ignore effects of windings etc.). Let us consider a simple example
to illustrate the point. We start with $T^2$ compactification of a bosonic
string such that we have nontrivial $G_{\alpha\beta}$ as initial background
to start with and
$B_{\alpha\beta}=0$. Thus under T-duality 
\bea
\label{texample}
M=\pmatrix{G^{11} & G^{12} & 0 & 0\cr
           G^{21} & G^{22} & 0 & 0\cr
            0    & 0  & G_{11} & G_{12} \cr
            0   & 0   & G_{21} & G_{22}\cr}   
            \rightarrow \pmatrix{G'^{-1} & -G'^{-1}B' \cr
                     B'G'^{-1} & G'-B'G'^{-1}B'\cr}
\eea
Explicitly,
\be
G'^{-1}=\pmatrix{G'^{11} & G'^{12} \cr
                 G^{21} &  G'^{22} \cr}~{\rm and}~
G'^{-1}B'=\pmatrix{G'^{11}B'_{11}+G'^{12}B'_{21} & G'^{11}B'_{12}+
G'^{12}B'_{22}\cr
G'^{21}B'_{11}+G'^{22}B'_{21} & G'^{21}B'_{12}+G'^{22}B'_{22}\cr}
\ee
where $G_{\alpha\beta}$ is symmetric, so is its inverse. Similarly, 
$G'_{\alpha\beta}$ is also symmetric, being the new metric of the internal space
and $B'_{\alpha\beta}$ is desired to be antisymmetric. Given the metric
$G'^{-1}$, we can extract $B'$.\\
We intend to consider a three point function and examine how T-duality
can be applied to generate a new three point function. Let us consider
a vertex with three massless scalar gravitons $G^{(1)}_{\alpha_1\beta_1}$,
$G^{(2)}_{\alpha_2\beta_2}$ and $G^{(3)}_{\alpha_3\beta_3}$. These arise
from compactification of the metric in $\hat D$-dimension to $D$-dimensions and
we may refer to  them as 'scalar graviton' from now on.
First we have to construct the vertex operators in the weak field approximation
as has been discussed earlier. The generic vertex operator for this case,
$h_{\alpha\beta}(X)\partial Y^{\alpha}{\bar\partial}Y^{\beta}$
is expressed as
\be
\label{vertexop}
 V^{(l)}_{\alpha_l\beta_l}=\epsilon^{h(l)}_{\alpha_l\beta_l}e^{iK_l.X_l}
\partial Y^{\alpha_i}{\bar\partial} Y^{\beta_l}
\ee
where $l=1,2,3$, $\epsilon^{h(l)}_{\alpha_l\beta_l}$ is the 
'polarization tensor' of the scalar graviton and it is 
 symmetric in its indices. Superscript $h(l)$ reminds that it is
'polarization tensor' associated with $h_{\alpha\beta}$ for the $l^{th}$ 
vertex operator. Note that $e^{iK_l.X_l}$ stands for the plane wave for
$l^{th}$ scalar and the plane wave depends only on the spacetime coordinates.
Thus, due to the absence of dependence on $(Y^{\alpha},~P_{\alpha})$ in the 
plane wave part, $h_{\alpha\beta}(X)$  propagates in the noncompact spacetime. 
We may call attention of  
the reader to the fact that here we are not taking into
account presence of winding modes etc. due to the choice of the propagating 
plane wave. \\
The vertex operator with a general structure is required to satisfy the 
following invariance properties \cite{wein1,wein2,phongd1,phongd2,ryu}.\\
(i) Spacetime translation. It is already satisfied since we have explicitly
factored out the 'plane wave' here.\\
(ii) Spacetime Lorentz transformation. Note that the indices $\alpha$ and 
$\beta$ represent those of compact coordinates. Thus Lorentz invariance is 
automatically respected.\\
(iii) General two dimensional coordinate transformation. We have already 
adopted the conformal gauge for worldsheet metric. \\
(iv) Conformal invariance of the vertex operator.\\
Notice that the requirements (iii) and (iv) are fulfilled if $k^2_l=0$ is
satisfied with the choice of the above vertex operator. \\
We shall not present the explicit evaluation of the three point vertex here
which can be computed by following standard procedures 
\cite{books,booksa,bookd,booke}
for example. Thus the three point function is
\be
\label{threept}
A_3= \bigg<:V^{(1)}(z_1,{\bar z}_1)::
 V^{(2)}(z_2,{\bar z}_2:: V^{(3)}(z_3{\bar z}_3): \bigg>
\ee
$z_i$ and ${\bar z}_i$ are complexified worldsheet coordinates. 
The normal ordered vertex operators are given by
\be
V^{(1)}=\epsilon^{h(1)}_{\alpha_1\beta_1}e^{iK_1.X}(z_1,{\bar z}_1)
\partial Y^{\alpha_1}(z_1){\bar\partial}Y^{\beta_1}({\bar z}_1),
\ee
\be
V^{(2)}=\epsilon^{h(2)}_{\alpha_2\beta_2}e^{iK_2.X}(z_2,{\bar z}_2)
\partial Y^{\alpha_2}(z_2){\bar\partial}Y^{\beta_2}({\bar z}_2),
\ee
\be
V^{(3)}=\epsilon^{h(3)}_{\alpha_3\beta_3}e^{iK_3.X}(z_3,{\bar z}_3)
\partial Y^{\alpha_3}(z_3){\bar\partial}Y^{\beta_3}({\bar z}_3),
\ee
We can arbitrarily fix three coordinates of the worldsheet $z_1, z_2, z_3$
for this  three point function  (\ref{threept}) and its computation is well
 known using techniques given in the text books 
 \cite{books,booksa,bookd,booke} 
and we shall not repeat them. However, we shall present how
we can generate a new three point function from $A_3$ by applying T-duality.
We could apply three different T-duality transformations on $V^{(1)}$,
$V^{(2)}$ and $V^{(3)}$; however, it will suffice to apply the transformation
on one of the vertex operators, say  $V^{(1)}$, to bring out the essential
argument. Given $h^{(1)}_{\alpha_1\beta_1}(X)$ now taken to be 
$\epsilon^{h(1)}_{\alpha_1\beta_1}e^{iK_1.X_1}$ we note that the plane wave part
i.e. $e^{iK_1.X_1}$ is inert under T-duality. Therefore, given a 'polarization'
$\epsilon^{h(1)}_{\alpha_1\beta_1}$ we can reconstruct 
$G_{\alpha_1\beta_1}=1+\epsilon^{h(1)}_{\alpha_1\beta_1}$ since 
$\epsilon^{h(1)}_{\alpha_1\beta_1}$ is an element of the  $\tilde M$-matrix.
We omit the explicit the factor of the plane wave $e^{iK_1.X_1}$ whose 
presence or absence does not affect the arguments in sequel.
After 
implementing the T-duality transformation we can reconstruct the new 
backgrounds; that is $\epsilon'^{h(1)}_{\alpha_1\beta_1}$ and generate another
component $\epsilon'^{b(1)}_{\alpha_1\beta_1}$ corresponding an antisymmetric
'polarization' tensor. Such a tensor would have appeared if we would 
initially start with backgrounds $G'_{\alpha_1\beta_1}$ and 
$B'_{\alpha_1\beta_1}$ and then resorted  to the weak field approximation and
eventually expressed the weak fields as product of a plane wave and 
'polarization' tensor. 
I shall present an explicit and simple $O(2,2)$ transformation matrix  
later and  another  example of a three point function which involves an excited
massive stringy state and two of the massless 'scalar gravitons'.\\
Let us consider three point function of a massive scalar arising from 
compactification of the first excited massive state of a closed string
and two scalars coming from the massless sector as above. We briefly
recall how such a scalar arises from the toroidal compactification. There
are three terms in the vertex operator for the first excited massive state.
For this case the one corresponding to the leading Regge trajectory is
a physical state, the other two can be gauged away. This argument becomes
physically transparent if we count physical degrees of freedom in the light 
cone gauge (also see \cite{wein1,wein2}).\\
The vertex operator is sum of four terms,
\be
\label{mvertex}
V_{(first)}=V_1+V_2+V_3+V_4
\ee
where
\be
V_1={\hat F}^{(1)}({\hat X})_{{\hat\mu\hat\nu},{\hat\mu} '{\hat\nu}'}
\partial{\hat X}^{\hat\mu}\partial {\hat X}^{\hat\nu}
{\hat X}{\bar\partial}{\hat X}^{\hat\mu}{\bar\partial}{\hat X}^{\hat\nu}
\ee
Note that ${\hat F}^{(1)}_{{\hat\mu\hat\nu},{\hat\mu} '{\hat\nu}'}$ is
symmetric under interchange of either ${\hat\mu}\leftrightarrow {\hat\nu}$
or ${\hat\mu}'\leftrightarrow {\hat\nu}'$ and also simultaneous interchange
of these pair of indices. However, there is no a priori reason to impose any
symmetry constraints on interchange of unprimed and primed indices.
The three other terms in (\ref{mvertex}) $V_2$, $V_3$ and $V_4$ are
\bea
\label{vertex2} 
 V_2= {\hat F}^{(2)}_{{\hat\mu\hat\nu},{\hat\mu}'}({\hat X})
\partial {\hat X}^{\hat\mu}\partial X^{\hat\nu}{\bar\partial}^2
{\hat X}^{{\hat\mu}'},~~~
 V_3= {\hat F}^{(3)}_{{\hat\mu},{\hat\mu}'{\hat\nu}'}({\hat X})
\partial ^2 {\hat X}^{\hat\mu}
{\bar\partial}{\hat X}^{{\hat\mu}'}{\bar\partial}{\hat X}^{{\hat \nu}'}
\eea
and
\be
V_4= {\hat F}^{(4)}_{{\hat\mu},{\hat \mu}'} \partial ^2 {\hat X}^{\hat\mu}
      {\bar\partial}^2
{\hat X}^{{\hat\mu}'}
\ee
We do not entertain  terms containing 
$ \partial{\bar\partial}{\hat X}^{\mu '}$ or 
${\bar\partial}\partial{\hat X}^{\hat\mu}$ in any vertex operators since they
vanish by virtue of free string equations of motion. This argument also
holds for vertex operators involving $X^{\mu}$ and compact coordinates
$Y^{\alpha}$ when we consider the case of compactified string coordinates.
The vertex operator $V_1$ corresponds to the state lying on leading 
Regge trajectory.
The requirement of the condition that $V_1$ be $(1,1)$ primary 
imposes constraints on 
${\hat F}^{(1)}({\hat X})_{{\hat\mu\hat\nu},{\hat\mu} '{\hat\nu}'}$.  
However, note
that same is not applicable to 
$ {\hat F}^{(2)}_{{\hat\mu\hat\nu},{\hat\mu}'}({\hat X})$,
${\hat F}^{(3)}_{{\hat\mu},{\hat\mu}'{\hat\nu}'}({\hat X})$ and
${\hat F}^{(4)}_{{\hat\mu},{\hat \mu}'}$. These three are not independently
$(1,1)$ primaries although they satisfy mass shell condition and appropriate 
\cite{ov1,ov1a,kr,ao}
transversality conditions. These three are related to ${\hat F}^{(1)}$. We have 
mentioned earlier that we shall only focus attentions on $V_1$ and therefore,
we shall not dwell on attributes of $V_2$, $V_3$ and $V_4$ in what follows.\\
If we demand 
${\hat F}^{(1)}({\hat X})_{{\hat\mu\hat\nu},{\hat\mu} '{\hat\nu}'}$ to be
of conformal weight $(1,1)$ with respect to $T_{++}({\hat X})$ and
$T_{--}({\hat X})$ the the following constraints are to be satisfied by
${\hat F}^{(1)}({\hat X})$
\bea
\label{shell}
({\hat\nabla }^2+2){\hat F}^{(1)}_{{\hat\mu}{\hat\nu} ,{\hat\mu}'{\hat\nu}'}
(\hat X)=0
\eea
where ${\hat \nabla}^2$ is the $\hat D$-dimensional Laplacian defined with
the flat space metric. Furthermore, in analogy with the graviton
${\hat h}_{\hat \mu\hat \nu}$ in $\hat D$-dimensional graviton, 
${\hat F}^{(1)}$ also satisfies transversality and gauge conditions
\cite{ov1,ov1a}.
\bea
\label{transg}
{{\hat F}^{(1){\hat\mu}}_{\hat\mu}},_{{\hat\mu}'{\hat\nu}'}+
2\partial^{\hat\mu}
\partial^{\hat\nu}{\hat F}^{(1)}_{{\hat\mu}{\hat\nu},{{\hat\mu}'}{{\hat\nu}'}}=0,~~~
{\rm and}~~~{{\hat F}^{(1){{\hat\mu}'}}_{{\hat\mu}{\hat\nu},{{\hat\mu}'} }}+
2\partial^{{\hat\mu}'}
\partial^{{\hat\nu}'}
{\hat F}^{(1)}_{{\hat\mu}{\hat\nu},{{\hat\mu}'}{{\hat\nu}'}}=0
\eea
Let us consider compactification of 
${\hat F}^{(1)}_{{\hat\mu}{\hat\nu} ,{\hat\mu}'{\hat\nu}'}(\hat X)$
to $D$ spacetime dimension. Thus the expression
for $V_1$ decomposes into sum of several terms and we discuss their structures
below very briefly.\\
(I) There will be a term where the $F^{(1)}$ tensor will have four 
$D$-dimensional Lorentz indices of the form
$F^{(1)}_{\mu\nu,\mu '\nu '}\partial X^{\mu}\partial X^{\nu}
{\bar\partial} X^{\mu '}{\bar\partial} X^{\nu '}$.\\

\noindent
(II) There will be a term will all internal indices i.e. those lie along
compact directions. This is of the form 
$F^{(1)}_{\alpha\beta,\alpha '\beta '}\partial Y^{\alpha}\partial Y^{\beta}
{\bar\partial} Y^{\alpha '}{\bar\partial} Y^{\beta '}$.\\
We remark that, so far as $F^{(1)}_{\mu\nu,\mu '\nu '}$ and 
$F^{(1)}_{\alpha\beta,\alpha '\beta '}$ are concerned, their symmetry 
properties under the interchange of indices are identical as those of 
${\hat F}^{(1)}_{{\hat\mu}{\hat\nu},{{\hat\mu}'}{{\hat\nu}'}}$ as noted 
earlier.\\

\noindent
(III) There will a structure with three Lorentz indices and one internal
with appropriate contractions with $\partial X$ and $\partial Y$ and/or
antiholomorphic parts. A generic term is 
$F^{(1)}_{\mu\nu,\mu '\beta '}\partial X^{\mu}\partial X^{\nu}
{\bar\partial} X^{\mu '}{\bar\partial} Y^{\beta '}$. There are three more
terms of this type with various indices and only feature is that there are
three Lorentz indices and one internal index (total four terms).\\

\noindent
(IV) There is another structure which has one Lorentz index and three
internal indices.\\
All the $F^{(1)}$'s satisfy the mass shell condition since ${\hat F}^{(1)}$
does so. Moreover, $~~$ 
$F^{(1)}_{\mu\beta,\alpha '\beta '}\partial X^{\mu}\partial Y^{\beta}
{\bar\partial} Y^{\alpha '}{\bar\partial} Y^{\beta '}$. There are also three
more terms of this form.\\

\noindent
(V) Finally there are six terms (in all) which have two Lorentz indices and two
internal indices. A generic term of this type is 
$F^{(1)}_{\mu\beta,\mu '\beta '}\partial X^{\mu} \partial Y^{\beta}
{\bar \partial} X^{\mu '}{\bar  \partial Y}^{\beta '}$.
We observe that the vertex operator defined in (I) above is inert under
T-duality. $F^{(1)}_{\mu\nu,\mu '\beta '}$, appearing in (III)
 transforms as a third rank tensor
and has only one internal index and therefore, the internal direction is
affected under T-duality transformation. Similarly, 
$F^{(1)}_{\mu\beta,\alpha '\beta '}$ transforms as a vector under Lorentz
transformations and therefore, it will satisfy transversality condition
like a gauge field. Finally, in case of the tensor appearing in (V) which
has two Lorentz indices and two internal indices, it transforms like
a second rank spacetime tensor. We have presented our arguments elsewhere,
how we can combine the 16 terms appearing in (I) to (V) to show that
this combined vertex operator can be cast in $O(d,d)$ invariant form. We
refer the interested reader to these articles \cite{jmmass,jmplb,revjm}.\\
Our focus is on the term defined in (II) which correspond to scalars
when we envisage their spacetime transformation properties. 
We recall that it is symmetric
under interchange of $\alpha\leftrightarrow \alpha '$ and also under
$\beta\leftrightarrow\beta '$. The states arise from composition of left and 
right mover and the form is consistent with the level matching condition
$L_0^Y-{\bar L}_0^Y=0$; the super script $Y$ refers to the fact that we are
considering only the compact coordinate sector. 
When we require the vertex operator
\be
\label{yvertex}
F^{(1)}_{\alpha\beta,\alpha '\beta '}(X)\partial Y^{\alpha}\partial Y^{\beta}
{\bar\partial} Y^{\alpha '}{\bar\partial} Y^{\beta '}
\ee
to be $(1,1)$ primary it satisfies two conditions
\be
\label{onshell}
(\nabla^2+2) F^{(1)}_{\alpha\beta,\alpha '\beta '}(X)=0, 
\ee
Here $\nabla^2$ is the $D$-dimensional flat space Laplacian. Moreover,
\bea
\label{gauge}
F^{(1){\alpha}}_{{\alpha},{\alpha}'{\beta}'}
=0,~~~
{\rm and}~~~ F^{(1){{\alpha}'}}_{{\alpha}{\beta},{\alpha}' }=0
\eea
If we compare this equation with (\ref{transg}) then we note that the second
term in each of the two relation present in (\ref{transg}) are absent. It is
understood easily since $ F^{(1)}_{\alpha\beta,\alpha '\beta '} $ depends only
on $X^{\mu}$. Thus any derivative with respect to $Y^{\alpha}$ acting on 
$F^{(1)}_{\alpha\beta,\alpha '\beta '}(X)$ vanishes. Now let us express
$F^{(1)}$ as a product of a 'polarization tensor and plane wave,
\be 
\label{planeyvertex}
F^{(1)}_{\alpha\beta,\alpha '\beta '}(X)=\epsilon_{\alpha\beta,\alpha'\beta'}
e^{iK.X}
\ee
It satisfies mass shell condition is $k^2=2$. Moreover,  
the polarization tensor has the properties that it is symmetric independently
under (i) $\alpha\leftrightarrow\beta$ and $\alpha'\leftrightarrow\beta'$ 
\be
\epsilon^{\alpha}_{\alpha\beta,\alpha'\beta'}=0,~~~ {\rm and} ~~~
\epsilon_{\alpha\beta,\alpha'\beta'}^{\alpha'}=0
\ee
We would like to construct the three point function for the massive scalar,
$k^2=2$, and two massless scalar gravitons coming from compactification of the 
$\hat D$-dimensional graviton.
\bea
\label{fvertex}
{\tilde A}_3= &&\epsilon_{\alpha\beta,\alpha'\beta'}
\epsilon^{h(1)}_{\alpha_1\beta_1}
\epsilon^{h(2)}_{\alpha_2\beta_2}\bigg<:e^{iK.X}\partial Y^{\alpha}
\partial Y^{\beta}{\bar\partial} Y^{\alpha'}{\bar\partial} 
Y^{\beta'}(z,{\bar z})
:\nonumber\\&&
:e^{iK_1.X_1}\partial Y^{\alpha_1}{\bar\partial} Y^{\beta_1}(z_1,{\bar z_1})::
e^{iK_1.X_2}\partial Y^{\alpha_2}{\bar\partial} Y^{\beta_2}(z_2,{\bar z}_2):
\bigg>
\eea
We may implement $T$-duality transformations on any one of the tree
vertex operators to generate new background as discussed before. \\
Let us consider a specific example when two spacial dimensions are compactified
on $T^2$. The duality group is $O(2,2)$ and it is isomorphic to 
$SL(2,R)\otimes SL(2,R)$. Note that the moduli $G_{\alpha\beta}$ and
$B_{\alpha\beta}$ parametrize the coset ${O(2,2)}\over{O(2)\otimes O(2)}$
and it is isomorphic to ${{SL(2,R)}\over{U(1)}}\otimes {{SL(2,R)}\over{U(1)}}$.
It is easy to see that former has 4 parameters and latter too has same number of
parameters. Therefore, an arbitrary $T$-duality transformation will be
specified by four parameters (say if we made two separate 
$SL(2,R)$ transformations).
We consider a very simple $O(2,2)$ transformation to illustrate our 
prescriptions. Let us start with diagonal $G_{\alpha\beta}=(g_{11},~g_{22})$
and set $B_{\alpha\beta} =0$ as the initial background configuration. Thus
the $M$-matrix is block diagonal. The $O(2,2)$ transformation is defined by
a single parameter as was adopted by us \cite{matrix2a}
to study the string cosmology in the 
Milne universe case and it has interesting consequences.  
Note that the choice of $\Omega$  is similar to the one adopted in  
\cite{matrix2a} with a slight difference due to the convention adopted here;
however, it fulfills all the requirements of an $O(d,d)$ matrix 
(\ref{transodd}).
\bea
\label{omega}
\Omega={1\over 2}\pmatrix{1+c & -s & c-1 & s\cr
               s & 1-c & s & 1+c \cr
               c-1 & -s  & 1+c & s \cr
               -s   & 1+c & -s & 1-c \cr}
\eea
where $ c=cosh\gamma$ and $ s= sinh\gamma$ and $\gamma$ is a bounded real
parameter. It is easy to check that $\Omega$ defined above (\ref{omega})
satisfies requirement of $O(2,2)$ transformation. The initial $M_{in}$-matrix
is
\bea
\label{initial}
M_{in}=\pmatrix{g^{11} & 0      & 0 & 0\cr
                 0  &    g^{22} & 0 & 0\cr
                 0 &     0 & g_{11} & 0\cr
                 0 &    0&  0 & g_{22} \cr }
\eea  
If we implement this duality transformation as stipulated in (\ref{transodd}),
we expect to generate new backgrounds. Let
\bea
\label{newm}
\Omega^T M_{in}\Omega=\pmatrix{U_{11}& U_{12} & U_{13} & U_{14} \cr
                              U_{21} & U_{22} & U_{23} & U_{24}  \cr
                              U_{31} & U_{32} & U_{33} & U_{34} \cr
                              U_{41} & U_{42} & U_{43} & U_{44} \cr }
\eea 
We can read off the transformed, $G'^{\alpha\beta}$ to be
\bea
\label{newmetric}
&&U_{11}=g'^{11}={1\over 4}\bigg[(1+c)^2g^{11}+s^2g^{22}+
(c-1)^2g_{11}+s^2g_{22}\bigg]
\nonumber\\&&
U_{12}=g'^{12}= {1\over 4}\bigg[s(1+c)g^{11}-s(1-c)g^{22}+
(c-1)sg_{11}+s(1+c)g_{22}
\bigg]\nonumber\\&&
U_{21}= g'^{21}= {1\over 4}\bigg[s(1+c)g^{11}-s(1-c)g^{22}+
s(c-1)g_{11}+s(1+c)g_{22}
\bigg]\nonumber\\&&
U_{22}=g'^{22}= {1\over 4}\bigg[s^2g^{11}+(1-c)^2g^{22}+s^2g_{11}+(1+c)^2g_{22}
\bigg]
\eea
We observe the appearance of 
the symmetric off diagonal elements $g'^{12}=g'^{21} $ as expected. We also 
note that already nontrivial $B'$ has been generated since we have nonzero 
elements $U_{13}, U_{14}, U_{23}$ and $U_{24}$ corresponding to $-G'^{-1}B'$.
After inverting $G'^{-1}$ we can extract $B'$. We present the matrix elements
for $G'^{-1}B'$ for sake of completeness.
\bea
\label{ginvb}
&& U_{13}={1\over 4}\bigg[(1+c)(c-1)g^{11}+s^2g^{22}+(c-1)(c+1)g_{11}+
          s^2g_{22} \bigg]\nonumber\\&&
U_{14}={1\over 4}\bigg[-s(1+c)g^{11}-s(1+c)g^{22}-s(c-1)g_{11}+
       s(1-c)g_{22}\bigg] \nonumber\\&&
U_{23}={1\over 4}\bigg[s(c-1)g^{11}-s(1-c)g^{22}+s(1+c)g_{11}+
       s(1+c)g_{22}\bigg]  \nonumber\\&&
U_{24}={1\over 4}\bigg[-s^2g^{11}+(1-c)(1+c)g^{22}-s^2g_{11}+
       (1+c)(1-c)g_{22}\bigg]
\eea
Given these matrix elements, we  can extract $B'$ since we can 
invert (\ref{newmetric}). A tedious and 
straight forward calculation shows $B'_{11}=B'_{22}=0$ as expected. Moreover,
$B'_{12}=-B'_{21}$. We give one matrix element below
\be
B'_{12}={\cal T}_1+{\cal T}_2
\ee
where
\bea
\label{term1}
{\cal T}_1=&&{s\over{16}}\bigg[-s^2(1+c)(g^{11})^2+s^2(1+c)(g^{22})^2+
s^2(1-c)(g_{11})^2-s^2(1+c)(g_{22})^2\nonumber \\&&
-2c(s^2+c^2+3)-2s^2cg^{22}g^{11}-2(c^3+c^2+c+2)g^{11}g_{22}+\nonumber \\&&
+2(1-c+c^2-c^3)g^{22}g_{11}-2s^2g_{22}g_{11} \bigg]
\eea
and
\bea
\label{term2}
{\cal T}_2=&&{{s^3}\over{16}}\bigg[ (1+c)(g^{11})^2+(c-1)(g^{22})^2+
(c-1)(g_{11})^2+(1+c)(g_{22})^2 \nonumber \\&&
+2\bigg(cg^{11}g^{22}+cg^{11}g_{22}+g^{22}g_{11}+(1+c)g_{11}g_{22}+2c\bigg)
\bigg]
\eea
An explicit calculation shows $B'_{12}=-B'_{21}$ and confirms that
$B'$ is antisymmetric.
The antisymmetric tensor background generated through the duality 
transformation in expressed in terms of $g_{11}, g_{22},g^{11}$ and
$ g_{22}$. In order to facilitate weak field expansions for the new 
backgrounds, our first step will be to adopt weak field expansion for the
initial metric, $g_{11} ~{\rm and}~ g_{22}$, its inverse 
$g^{11} ~{\rm and}~ g^{22}$,
\be
g_{\alpha\beta}=\delta_{\alpha\beta}+h_{\alpha\beta},
~~g^{\alpha\beta}=\delta^{\alpha\beta}+h^{\alpha\beta},~~
h^{\alpha\beta}=-\delta^{\alpha\alpha '}\delta^{\beta\beta '}h_{\alpha '\beta '}
\ee
and $ \alpha,\alpha, '\beta,\beta ' =1,2$. In the expression
for the new inverse metric (\ref{newmetric}), we find that off diagonal elements
have appeared and these matrix elements will need to be expanded in the weak
field expansion of original $g_{\alpha\beta}$ and  $g^{\alpha\beta}$. 
Another point is that, when we resort to weak field expansions
for original metric, the newly generated metric starts with a constant 
$(\ne 1)$. Thus we should scale this constant part to have a weak field
expansion, similar to the initial metric. Second point worth mentioning is that
to obtain weak field expansion for $B'$-field, we should expand the initial
metric in this approximation. We find that after this expansion, a constant
piece appears in nonzero components of $B'$ as can be seen from the 
expressions for ${\cal T}_1$ and ${\cal T}_2$ given by (\ref{term1}) and
(\ref{term2}). However, as we argued earlier, the
constant piece does not contribute to the equations of motion. Therefore, we may
ignore this constant piece, if we so desire, in our subsequent discussions. The 
antisymmetric tensor field continues to maintain its antisymmetry property
after the expansion. Now if we express 
$h_{\alpha\beta}(X)=\epsilon^{h}_{\alpha\beta}e^{iK.X}$ 
then it is easy to see that
\be
\label{t2trans}
\epsilon^{h}_{\alpha\beta}\rightarrow \epsilon'^{(h')}_{\alpha\beta}+
\epsilon'^{(b')}_{\alpha\beta}
\ee
We can derive this relation starting from plane wave approximation for 
$h_{\alpha\beta}$, then construct ${\tilde M}_{in}$. After we implement the
transformation (\ref{omega}) we obtain $M_{f}$. Next we derive 
${\tilde M}_{f}$ and then extract $\epsilon'^{(h')}$ and $\epsilon'^{(b')}$.
Therefore, if we applied this T-duality transformation on one of the legs of the
three point function (\ref{threept}) we shall generate a new three point 
function with $B'$-background in the weak field approximation along with the
$G'$. Same argument
holds for the other legs, consisting of 'scalar gravitons' in the 
three point amplitude, ${\tilde A}_3$ given by (\ref{fvertex}). \\
If we want to apply the T-duality transformation on the massive scalar
appearing in (\ref{fvertex}), it is not so straight forward as the case of
the 'scalar graviton'. In order to examine how it works, let us explicitly
write the terms for this scalar vertex operator (\ref{yvertex}). It will have
altogether 16 terms. We mention in passing that all the tensorial 
indices corresponding to compact directions are raised and lowered 
by Kronecker delta here and in particular 
$P_{\alpha}=\delta _{\alpha\beta}{\dot Y}^{\beta}$. 
We can group the 16 terms into
five classes and define each term to be a vertex function. Thus the vertex
operator (\ref{yvertex}) is sum of 16 vertex functions. I list them below for
later conveniences:\\
{\bf 1.} A term which has products of 
all $\{ P_{\alpha}\}$'s suitably contracted.  
We have dropped superscript $(1)$ of the $F$ and the expression is :
\be
F^{\alpha\beta,\alpha'\beta'}P_{\alpha}P_{\beta}P_{\alpha'}P_{\beta'}
\ee

\noindent
{\bf 2.} Another term which has products of all $\{Y^{\alpha} \}$'s. It is
\be
F_{\alpha\beta,\alpha'\beta'}Y'^{\alpha} Y'^{\beta}Y'^{\alpha'}Y'^{\beta'}
\ee

\noindent
{\bf 3.} There are four terms which are composed of product of three 
$P_{\alpha}$ 
and
one $Y^{\beta}$:
\be
\bigg[-F^{\alpha\beta,\alpha'}_{\beta'}P_{\alpha}P_{\beta}P_{\alpha'}
Y'^{\beta'} -F^{\alpha\beta,\beta'}_{\alpha'}P_{\alpha}P_{\beta}P_{\beta'}
Y'^{\alpha'}+ F^{\alpha,\alpha'\beta'}_{\beta}P_{\alpha}P_{\alpha'}P_{\beta'}
Y'^{\beta}+  F^{\beta,\alpha'\beta'}_{\alpha}P_{\beta}P_{\alpha'}P_{\beta'}
Y'^{\alpha} \bigg]
\ee

\noindent
{\bf 4.} There are another group of terms which have product of 
one $P_{\alpha}$ 
and three $\{Y'^{\beta} \}$. These are
\be  
\bigg[
+F^{\alpha}_{\beta,\alpha'\beta'}P_{\alpha}Y'^{\beta}Y'^{\alpha'}Y'^{\beta'}
+F^{\beta}_{\alpha,\alpha'\beta'}Y'^{\alpha}P_{\beta}Y'^{\alpha'}Y'^{\beta'}
-F^{\alpha'}_{\alpha\beta,\beta'}Y'^{\alpha}Y'^{\beta}P_{\alpha'}Y'^{\beta'}
-F^{\beta'}_{\alpha\beta,\alpha'}Y'^{\alpha}Y'^{\beta}Y'^{\alpha'} P_{\beta'}
 \bigg]
\ee
\noindent
{\bf 5.} There are six terms in this class. Each term is a product of a 
pair of
$P_{\alpha}$'s and a pair of $Y'^{\beta}$.
\bea
&&
+F^{\alpha\beta}_{\alpha'\beta'}P_{\alpha}P_{\beta}Y'^{\alpha'}Y'^{\beta'}
-F^{\alpha,\alpha'}_{\beta,\beta'}P_{\alpha}Y'^{\beta}P_{\alpha'}Y'^{\beta'}
-F^{\alpha,\beta'}_{\beta,\alpha'}P_{\alpha}Y'^{\beta}Y'^{\alpha'}P_{\beta'}
  \nonumber \\ && 
-F^{\beta,\alpha'}_{\alpha,\beta'}Y'^{\alpha}P_{\beta}P_{\alpha'}Y'^{\beta'}
-F^{\beta,\beta'}_{\alpha,\alpha'}Y'^{\alpha}P_{\beta}Y'^{\alpha'}P_{\beta'}
+F^{\alpha',\beta'}_{\alpha,\beta}Y'^{\alpha}Y'^{\beta}P_{\alpha'}P_{\beta'}
 \eea
Let us discuss the T-duality, $P_{\alpha}\leftrightarrow Y'^{\alpha}$, which
is the discrete $Z_2$ duality. We observe that the two terms given in
${\bf (1)}$ and ${\bf (2)}$ above interchange. 
Therefore, if the $F$'s interchange accordingly
they are $Z_2$ symmetric. Next, consider the four terms in ${\bf (3)}$ and
four terms in ${\bf (4)}$. Under this duality we find the combined eight terms
are also T-duality invariant. Finally the six terms in ${\bf (5)}$ transform
among themselves under $P_{\alpha}$ and $Y'^{\alpha}$ interchange. Furthermore,
I have proposed a prescription to express vertex operators of excited massive
levels in T-duality invariant form. Therefore, if we start with any vertex
function given above then by a suitable $O(d,d)$ transformation or series of
such transformations, we can generate other terms \cite{jmmass}. 
For example we can
write $P_{\alpha}$ as
\be
\label{projecto}
\pmatrix{P_{\alpha}\cr 0\cr} ={\cal P}_+\pmatrix{P_{\alpha} \cr
                            Y'^{\alpha} \cr},~
{\cal P}_+=\pmatrix{1 & 0 \cr
                    0 & 0 \cr},~
\pmatrix{0\cr Y'^{\alpha}}={\cal P}_-\pmatrix{P_{\alpha} \cr Y'^{\alpha} \cr}
{\cal P}_-=\pmatrix{0 & 0\cr 0 & 1 \cr}
\ee
as mentioned earlier the double $P_{\alpha} $ and  $Y'^{\alpha}$ transform like 
$O(d,d)$ vector. Thus ${\cal P}_+$ projects out the upper component. Similarly,
 projection operator  ${\cal P}_-$  projects out the lower component
$Y'^{\alpha}$  (see \cite{jmmass} for details). Moreover,
the lower component  $Y'^{\alpha}$ can be flipped if we operate the  
$ O(d,d)$ metric $\eta$ on it. \\
In view of the above arguments, we can start with three point function,
${\tilde A}_3$, of the excited first scalar massive level with two 
'scalar graviton' in a given configuration for the massive state 
and implement $O(d,d)$ transformation to generate another configuration.
Since T-duality symmetry is expected to be valid in each order of perturbation
theory my conjecture is that these transformations to generate new 
three point functions are expected to hold good beyond the tree level,
order by order in perturbation theory.\\
We present another argument which makes T-duality transformation 
intuitively more appealing for at least some special states when we consider
the tree level amplitudes. These are
the massive scalars 
 coming from compactification of the state in ${\hat D}$-dimension which
lie on the leading Regge trajectory. We invoke the arguments of Kawai,
Llewellen and Tye \cite{klt} who provide a very elegant technique to establish
relationship between open string and closed string scattering amplitudes
at the tree level. First we recapitulate their arguments very briefly and then
apply them to the case at hand. In particular, we exploit the formulation of
KLT for construction of vertex operators of excited closed string states to
the end that the theory is compactified to lower dimension. Subsequently.
we modify their approach appropriately to propose vertex operators which 
have desirable transformation properties under the $T$-duality transformation.
The physical degrees of freedom, for a free
closed string, are composed of left-movers and right-movers since the
oscillators from these two sectors act on the vacuum. We can treat them 
independently. Second point to note is that a closed string with only
left-moving or right-moving mode is intimately related to open string theory
in the mathematical sense. When we compute closed string tree level 
scattering amplitudes, we have product of vertex operators with multiple 
integrals. The integrand can be expressed in a factorized form into left moving
and right moving sectors. We can identify that these are same as the ones
appearing in the open string amplitudes. Of course, there are some subtleties
in expressing the N-point closed string tree amplitudes as products of N-point
open string amplitudes at tree level. In particular the three point and four
point closed string amplitudes can be explicitly written in the factorized
form for products of three point and four point open string amplitudes
respectively. We recall: (a) the closed string point point function
$A^{(3)}_{\rm closed}$ can be determined by fixing the three worldsheet variable
arbitrarily (the so called three Koba-Nielsen variables). As KLT \cite{klt}
demonstrated, this three point function is determined in terms of the product
of two open string three point function. (b) In case of the 4-point function,
the closed string amplitude can be computed once we fix three variables and
an analogous factorization was demonstrated. Moreover, they have laid down
the prescriptions to compute N-point closed string tree amplitude in the
factorized form of products of tree level open string amplitudes. The prospect
of going beyond tree level in this kind of approach has been explored in
\cite{mp}.\\
Our principal goal is not so much as to utilize the KLT factorization
formula to compute scattering amplitudes; rather we take the clue from this 
work to construct vertex operators for excited stringy states and study their 
duality properties from the perspectives of KLT. We argue that their 
formulation  simplifies some calculations for the states coming from the
leading Regge trajectories when we envisage the corresponding vertex operators.
In order to achieve this objective we recapitulate how vertex operators
associated with the excited states be constructed starting from the tachyon
vertex operator. While employing the Green's function technique (or the
path integral approach) we should use the appropriate boundary conditions for
the open string coordinates and closed string coordinates. We shall not dwell
on them here. Let us begin with the open string tachyon vertex operator:
$V(k, X)=:e^{ik.X}:$. The N-point amplitude is product of these vertex
operators located at N different points and integrated over appropriate
worldsheet variable. The essential point to note is that the emission vertex
for the first  excited state can be extracted by looking at OPE of the 
product of two tachyon vertex operators. From a simple analysis KLT arrive at
the known vector boson vertex operator
\be
\label{vectb}
V(\zeta,k,X)=\zeta^{\mu}:\partial X^{\mu}e^{ik.X}:
\ee
Their observation that this vertex operator can be extracted  from a 
modified expression of tachyon vertex operator leads to interesting and
important consequences for the results of \cite{klt}. The vertex
operator they propose is
\be
\label{klt1}
V(\zeta,k,X)=:e^{ik.X+i\zeta^{\mu}\partial X^{\mu}}:
\ee
The vertex operator (\ref{vectb}) can be obtained if we expand (\ref{klt1})
in powers of $\zeta^{\mu}$ and retain only the linear term.
Thus, with this trick, the N-point scattering amplitude of massless vector
bosons of open string can be computed.  Moreover, they propose that the 
vertex operators for excited massive open string states can also be constructed
\be
\label{exopen}
V(\zeta,k,X)= \zeta_{\mu_1\mu_2..\mu_n}:\partial X^{\mu_1}\partial X^{\mu_2}..
\partial X^{\mu_n}e^{ik.X}:
\rightarrow :e^{i(k.X+\zeta^{(1)}.\partial X+...\zeta^{(m)}.\partial X)}:
\ee
where $\zeta^{(m)}.\partial X =\zeta^{(m)}_{\mu}.\partial X^{\mu}$. Thus
we can compute amplitudes by assembling multilinear terms in 
$\zeta^{(l){\mu}}$ after evaluating products of N such operators as argued
by KLT. We can  resort to similar formulation for evaluation of the
amplitudes for closed string excited states. Let us consider the graviton
vertex
 in this perspective. Moreover, we intend to explore duality properties
of the vertex operator in this formalism when we compactify the theory. 
Therefore, we start with vertex
operator in $\hat D$-dimensions. In our notation, if we express 
(\ref{oldgvertex}) in terms of the polarization tensor and plane wave 
\be
\label{newgvertex}
{\hat V}({\hat\epsilon},{\hat k},{\hat X})=
{\hat \epsilon}_{\hat\mu\hat\nu'}:\partial {\hat X}^{\hat \mu}
{\bar \partial} {\hat X}^{{\hat\nu}'} e^{i{\hat k}.{\hat X}}:
\ee 
This vertex operator satisfies the mass shell condition  and
the transversality conditions respectively 
\be
\label{new11}
{\hat k}^2=0,~~{\rm and}~~~
{\hat k}^{\hat\mu}{\hat\epsilon}_{\mu\nu'}=
{\hat k}^{\hat\nu'}{\hat\epsilon}_{\mu\nu'}=
\ee
In the formalism of KLT, we can express the vertex operator as
\be
\label{kltgvertex}
{\hat {\tilde V}}({\hat\zeta},{\hat{\bar\zeta}},{\hat k},{\hat X})=:e^{ik.X+
i{\hat\zeta}.\partial {\hat X}+i{\hat{\bar\zeta}}.{\bar\partial} {\hat X}}:
\ee
Here ${\hat\zeta}_{\hat\mu}$ and ${\hat{\bar\zeta}}_{\nu'}$ two vectors 
contracting with $\partial {\hat X}^{\hat \mu}$ and 
${\bar \partial} {\hat X}^{{\hat\nu}'}$ respectively.
In order to establish relations between (\ref{newgvertex})  and 
(\ref{kltgvertex}) we should expand the latter and keep terms proportional to
 ${\hat\zeta}_{\hat\mu}{\hat{\bar\zeta}}_{\nu'}$.      
Note that equations (\ref{newgvertex}) and (\ref{kltgvertex}) differ by a
factor of $'i'$ it can be easily taken care of by a suitable redefinition. 
Moreover, ${\hat \epsilon}_{\hat\mu\hat\nu'}$ is symmetric 
in its indices. One more point might be worth mentioning here. The vertex
operator extracted from (\ref{kltgvertex}) as the coefficient of the 
biliear in  ${\hat\zeta}_{\hat\mu}$ and ${\hat{\bar\zeta}}_{\nu'}$ is to be
understood as the weak field expression for graviton vertex operator.   Thus to
bring out this aspect explicitly, we could reintroduce the small parameter
'$a$' through a scaling
\be
{\hat\zeta}_{\hat\mu}\rightarrow {\sqrt a} {\hat\zeta}_{\hat\mu},
~~{\rm and}~~ {\hat{\bar\zeta}}_{\nu'}\rightarrow{\hat{\bar\zeta}}_{\nu'}
\ee
Note, however,  that
the product ${\hat\zeta}_{\hat\mu}{\hat{\bar\zeta}}_{\nu'}$ can be decomposed
as sum of a symmetric and antisymmetric second rank tensor. The symmetric part
is identified with the graviton polarization tensor whereas the antisymmetric
one is that of the ${\hat B}_{\hat\mu{\hat\nu}'}$. Both the polarization
tensors satisfy the transversality constraints. \\
Let us examine how the compactification scheme will be introduced in this
frame work. As we have explained earlier the polarization tensors will
be decomposed to four pieces in this case i.e.
${\hat\epsilon}_{{\hat\mu}{\hat\nu'}}
\rightarrow \{\epsilon_{\mu\nu'}, \epsilon_{\alpha\nu'}\epsilon_{\mu\beta'}
~{\rm and}~ \epsilon_{\alpha\beta'} \}$. If we look at the symmetric part of
the product 
${\hat\zeta}_{{\hat\mu}}{\hat{\bar\zeta}}_{{\hat\nu}'}$ it will decompose
into $\{ \zeta_{\mu}{\bar\zeta}_{\nu}, \zeta_{\mu}{\bar\zeta}_{\beta'}, 
\zeta_{\alpha}{\bar\zeta}_{\nu'}~{\rm and}~ 
\zeta_{\alpha}{\bar\zeta}_{\beta'} \}$. 
Obviously, the first one corresponds to polarization tensor of $D$-dimensional
graviton, the next two are polarization vectors of gauge bosons and last one
is to be identified with 'scalar graviton'. Let us carefully analyze the
terms appearing in the exponential of (\ref{kltgvertex}). We have argued 
earlier that plane wave part of the vertex operator 
does not depend on the compact coordinates; thus ${\hat k}.{\hat X}
\rightarrow k.X$. Then the pair of  terms 
$\{ {\hat\zeta}.\partial {\hat X},~{\hat{\bar\zeta}}.{\bar\partial} {\hat X} \}
\rightarrow \{ \zeta_{\mu}\partial X^{\mu}+\zeta_{\alpha}\partial Y^{\alpha},
{\bar\zeta}_{\nu'}{\bar\partial}X^{\nu'}+{\bar\zeta}_{\beta'}{\bar\partial}
Y^{\beta'} \} $. Therefore, when we expand the exponential and retain the
bilinear terms as alluded to above, we can identify the symmetric part
of the 'polarization' tensors to be the one which couples to the
'scalar graviton' after we have compactified the theory to lower dimension.
So far we have not addressed the issue of duality in this scheme. Now I 
propose the following for the massless sector (the first excited states)
and this argument can be generalized to all excited massive levels lying
on the leading Regge trajectory. 
In order to express the compactified vertex operators in KLT scheme we shall
go through following steps (I shall confine to the products of
$\partial Y^{\alpha}{\bar\partial}Y^{\alpha'}$ here)\\
1. We must express $\partial Y=P+Y'$ and ${\bar\partial}Y=P-Y'$ in terms of
$O(d,d)$ vectors \cite{jmmass}. This is achieved by introducing projection
operators given by (\ref{projecto}). If we define an $O(d,d)$ vector
\be
{\cal W} =\pmatrix{P \cr Y'\cr}
\ee
then we can express 
\bea
\label{py}
P+Y'={1\over 2}\bigg({\bf P}_+{\cal W}+ {\bf\eta}{\bf P}_-{\cal W}\bigg),~~
P-Y'={1\over 2}\bigg({\bf P}_+{\cal W}- {\bf\eta}{\bf P}_-{\cal W}\bigg)
\eea
In the above equation (\ref{py}) $P\pm Y'$ are to be understood as $O(d,d)$
column vectors. We remind 
 that $\bf\eta$ flips lower component $Y'$ vector to an upper component
one. In order to construct an $ O(d,d)$ invariant object out of
$\zeta_{\alpha}\partial Y^{\alpha}$ and 
${\bar\zeta}_{\alpha}{\bar\partial} Y^{\alpha}$, we introduce doublets,
which transform as $O(d,d)$ vectors, as
\bea
\label{doublezeta}
\zeta\rightarrow\pmatrix{\zeta \cr {\tilde \zeta}\cr}={\bf Z},~~{\rm and}~~
{\bar\zeta}\rightarrow\pmatrix{{\bar\zeta}\cr{\tilde{\bar\zeta}}}={\bf{\bar Z}}
\eea
Thus if we demand that the inner product
\bea
\label{inv1}
{1\over 2} {\bf Z}^T.\bigg({\bf P}_+{\cal W}+ {\bf\eta}{\bf P}_-{\cal W}\bigg),
~~{\rm and}~~
{1\over 2} {\bf{\bar Z}}^T.\bigg({\bf P}_+{\cal W}- {\bf\eta}{\bf P}_-
{\cal W}\bigg)
\eea
be $O(d,d)$ invariant, the compactified sector of KLT vertex can be cast in 
$O(d,d)$ invariant form since the plane wave part $e^{ik.X}$ is inert 
under T-duality transformations. The introduction the doublet of 
polarization tensors should not surprise us. We encounter appearance of
extra fields/parameters in a Lagrangian in order to enforce manifest invariance
of the Lagrangian. In fact, when we expressed the Hamiltonian density in
manifestly $O(d,d)$ invariant form (\ref{hamiltonian}), we introduced 
$2d \times 2d$ $M$-matrix which is of larger dimension. At that point we knew
that the standard form of the canonical Hamiltonian can be expressed in terms
of $G_{\alpha\beta}$ and  $B_{\alpha\beta}$ which together have only $d^2$
independent elements. Thus $M$ is not an arbitrary $O(d,d)$ matrix
and it is a special one (see Section 3 of ref.\cite{ms} for details).
We can write analog of (\ref{kltgvertex})
in our formulation for the first excited massless scalar graviton state
\bea
\label{dualityvert}
V_{\rm scalar}({\bf Z},{\bf{\bar Z}},X)=:e^{ik.X+{1\over 2}
i{\bf Z}^T({\bf P}_+{\cal W}+ {\bf\eta}{\bf P}_-{\cal W})+{1\over 2}
i{\bf{\bar Z}}^T ({\bf P}_+{\cal W}- {\bf\eta}{\bf P}_-{\cal W})}:
\eea
We can extract the scalar graviton vertex by picking up the bilinear term
in ${\bf Z}$ and ${\bf{\bar Z}}$. This prescription to construct vertex
operators corresponding to compact directions is more elegant and simple than
my initial proposal \cite{jmmass}. However, we note that we have succeeded
in utilizing KLT technique for states lying on the leading Regge trajectory
in $\hat D$-dimension and then focusing on the scalars arising from 
compactification to $D$-dimension. It is clear that the same argument will hold
for background vector fields as well. 
We can also adopt the prescriptions of \cite{klt} for
excited massive levels coming from compactification of corresponding states in
$\hat D$-dimension to lower dimension. \\
For excited massive levels in $\hat D$-dimensions one is expected to introduce
a number of polarization tensors ${\hat\zeta}^{(1)}_{{\hat\mu}_1},
{\hat\zeta}^{(2)}_{{\hat\mu}_2},.... {\hat\zeta}^{(n)}_{{\hat\mu}_n}$ and
correspondingly  
${\hat{\bar\zeta}}^{(1)}_{{\hat\mu'}_1},....
{\hat{\bar\zeta}}^{(n)}_{{\hat\mu'}_n}$ in the exponential. Then expand the
exponential and keep multilinear terms, however, the number of
${\hat\zeta}^{(i)}_{{\hat\mu}_i}$ should be exactly the same as those  of 
${\hat{\bar\zeta}}^{(i)}_{{\hat\mu'}_i}$ in order to fulfill level matching
condition for the closed string. When we compactify to lower dimensions, we
shall have states with Lorentz tensor indices and internal indices as has been
pointed out earlier. 
Therefore, as before, we should introduce doublets of $\zeta$ and 
${\bar\zeta}$ when we have compactified to lower dimensions i.e.
we have a set of doublets
\bea
\label{nvect} 
{\zeta^{(l)}_{\mu_l}}\rightarrow\pmatrix{\zeta^{(l)}\cr {\tilde\zeta}^{(l)}\cr},
~~{\rm and}~~{\bar{\zeta}}^{(l')}_{\mu_{l'}}\rightarrow
\pmatrix{{\bar\zeta}^{(l')}\cr {\bar{\tilde\zeta}}^{(l')}\cr},
\eea
Note that $l=1,2..n$ and $l'=1,2...n$ in order to satisfy level matching
condition.
\\
We are in a position to discuss the T-duality transformation on the vertex
operator associated with excited massive scalar in this frame work. Let us
consider the three point function (\ref{fvertex}). The vertex operator of
the massive state (\ref{planeyvertex}) can be re-expressed as follows
\bea
\label{mvertex}
F^{(1)}= &&:{\bf Z}_i{\bf Z}_j{\bar{\bf Z}}_{i'}{\bar{\bf Z}}_{j'} e^{ik.X}
({\bf P}_+{\cal W}+ {\bf\eta}{\bf P}_-{\cal W})_i
({\bf P}_+{\cal W}+ {\bf\eta}{\bf P}_-{\cal W})_j \nonumber \\&&
({\bf P}_+{\cal W}- {\bf\eta}{\bf P}_-{\cal W})_{i'}
({\bf P}_+{\cal W}- {\bf\eta}{\bf P}_-{\cal W})_{j'}:
\eea
The inspection of (\ref{mvertex}) shows that the contraction of indices
 of ${\bf Z}_i{\bf Z}_j$ correspond to the recasting of terms 
$\partial Y^{\alpha}\partial Y^{\beta}$ in terms of the projections of
${\cal W}$'s whereas  indices of ${\bar{\bf Z}}_{i'}{\bar{\bf Z}}_{j'}$
contract with those coming from recasting of 
${\bar\partial} Y^{\alpha'}{\bar\partial} Y^{\beta'}$. Therefore, the product
is symmetric under $i\leftrightarrow j$ and  $i'\leftrightarrow j'$. However,
there is no compelling reason regarding such properties under
$i\leftrightarrow i'$ and $j\leftrightarrow j'$. We remind the reader about
construction of graviton vertex operator in the KLT formalism: the symmetric
product of $\zeta_{\mu}\zeta_{\nu'}$ was associated with graviton polarization
whereas the antisymmetric product with the polarization tensor of $B_{\mu\nu}$.
Moreover, both these backgrounds couple to 
$\partial X^{\mu}{\bar\partial} X^{\nu'}$. For the case at hand the product
like, say, ${\bf Z}_i{\bar{\bf Z}}_{j'}$ will have symmetric and antisymmetric
decompositions in general.\\
We can implement suitable $O(d,d)$ transformations on ${\bf Z}_i,{\bf Z}_j,
{\bar{\bf Z}}_{i'}~~ {\rm and }~~{\bar{\bf Z}}_{j'}$ each one of which 
transform as component of an $O(d,d)$ vector (doublet). Therefore, if we
start with a given three point function with a specific form of 
(\ref{dualityvert}), we can generate a new three point function keeping the
two scalar graviton legs unaltered if we so desire. This is an example
how T-duality transformation can act with two scalar graviton and a massive
scalar $k^2=2$. The procedure to construct vertex operators for excited
states of a compactified closed string has been outlined above. We can 
implement T-duality transformations on amplitudes involving an arbitrary
vertex operator of this type to generate another amplitude.\\
Now I consider a simple four point amplitude and discuss how T-duality
transformation can be applied in the context of the present formulation.
Let us consider a four point function for scattering  a gauge boson and
a tachyon in $D$-dimensions. We assume that the gauge boson, denoted 
as $A^{\alpha}_{\mu}(X)$ owes its origin
to compactification of $\hat D$-dimensional graviton. The vertex operator is
\be
\label{gaugevertexo}
V_{A}(\epsilon,k,X)=\epsilon_{\mu\alpha}:\partial X^{\mu} e^{ik.X}
{\bar\partial} Y^{\alpha}
\ee
Here $\epsilon_{\mu\alpha}$ is the polarization vector satisfying 
transversality condition  $k^{\mu}\epsilon_{\mu\alpha}=0$. If we were to
generate this vertex operator in the KLT formulation, we shall follow the
prescription
\bea
\label{vectklt}
V_{A}^{\rm klt}=:e^{ik.X+i\zeta_{\mu}\partial X^{\mu}
+i{\bar\zeta}_{\alpha}{\bar\partial} Y^{\alpha}}:
\eea
then keep the bilinear term $\zeta_{\mu}{\bar\zeta}_{\alpha}$ in the expansion.
This term is to be identified with gauge boson polarization tensor.
\\
Let us consider the four point amplitude amplitude for 
$T+A_{\mu\alpha}\rightarrow T'+A'_{\nu\beta}$ where $T ~{\rm and}~  T'$ are the 
incoming and outgoing  tachyons
and $A_{\mu\alpha}~{\rm and}~ A'_{\nu\beta}$ are incoming and outgoing
gauge bosons which arise due to compactification of the $\hat D$-dimensional
graviton to lower dimensions. In fact from the definition of the vierbein they
should have indices $\{ A^{\alpha}_{\mu},A^{\beta}_{\nu} \}$; 
recall, however,  that
the internal indices are raised and lowered by 
$\delta^{\alpha\beta}$ and $\delta_{\alpha\beta}$ in the weak field 
approximation. The amplitude is
\bea
\label{fourpt}
A_{4}=\epsilon_{\mu\alpha}\epsilon_{\nu\beta} \bigg<&&:e^{ik.X}(z_1,{\bar z}_1):
:\partial X^{\mu} e^{ik.X}\partial Y^{\alpha}(z_2,{\bar z}_2) :
\nonumber \\&&
:e^{ik.X}(z_3,{\bar z}_3):
:\partial X^{\nu} e^{ik.X}{\bar\partial} Y^{\beta}(z_4,{\bar z}_4) :\bigg>
\eea
where $\epsilon_{\mu\alpha}\epsilon_{\nu\beta}$  are vector polarization 
tensors of the two gauge bosons (includes internal index as well).
These   can be decomposed as
\be
\label{decomp}
\epsilon_{\mu\alpha}=\zeta_{\mu} {\bar\zeta}_{\alpha}, ~~~{\rm and}~~~
\epsilon_{\nu\alpha}=\zeta_{\nu} {\bar\zeta}_{\beta}
\ee
when we adopt the KLT formalism.
We intend to explore implications of $T$-duality on this amplitude 
(\ref{fourpt}). The two tachyon vertex operators are unaffected by T-duality
transformation as argued earlier because these are plane waves $e^{ik.X}$.
We also know how to convert ${\bar\partial} Y^{\alpha}$ in the exponential
appearing in (\ref{vectklt}) to $O(d,d)$ vector. A careful inspection
shows that the internal vector ${\bar\zeta}_{\alpha}$ needs an other
partner and presence of this can be justified from an old result \cite{ms}.
It was shown by us that the $d$ vector potentials, associated with the
isometries, which arise from compactification of graviton, combine with another
$d$ vector potentials coming from compactification of $\hat B$-field to 
constitute a double which transform as $O(d,d)$ vector as $\cal W$ is also such
a vector. To be more precise, $A^{(1)\alpha}_{mu}$ and
$A^{(2)}_{\mu\alpha}={\hat B}_{\mu\alpha}+B_{\alpha\beta}A^{(1)\beta}_{\mu}$ 
form an $O(d,d)$ and these together have $2d$ components which transform as
$O(d,d)$ vector. However, our argument that follows, is not affected  even
if we incorporate these subtleties. 
Therefore, incorporating the above argument we see that an 
amplitude describing scattering of a tachyon with a gravi-photon under the
T-duality transformation will give us an amplitude which describes
 scattering of gravi-photon on a tachyon going to tachyon and 
admixture of gravi-photon
and axi-photon in the final state. 
The gauge boson coming from compactification of
$\hat B$-field is termed as axi-photon. It is important to recall that Sen
\cite{senoddlr,hassen} proposed $O(d)\otimes O(d)$ transformation to 
generate new 
backgrounds from initial ones in the context of black holes. In fact his
proposal can be implemented very elegantly when we look at the KLT prescription
for strings with compact dimensions since one transformation is $O(d)_L$ and
the other one is  $O(d)_R$.\\
To complete this section, we note that we have set up formalism for 
the three point functions involving three scalar gravitons as well as one
for one excited massive (scalar) state and two scalar gravitons in the weak
field approximation.  The T-duality can be implemented on scalar gravitons
utilizing our formalism. We have shown that the modified version of KLT 
formalism has certain advantage over my earlier prescriptions. Moreover,
the new formulation is better suited for studying $T$-duality
transformation on appropriate scattering amplitudes.

\section{Summary and Discussions}

\noindent 
Our goal in this paper is to study scattering of close string states 
which arise due to compactification to lower spacetime dimension and explore
the consequences of T-duality symmetry. In order to envisage scattering in
string theory, in the first quantized formulation,
 we have to adopt an approximation scheme. The accepted procedure
is to construct vertex operators which are $(1,0)$ and $(0,1)$ primary with 
respect to $T_{++}$ and $T_{--}$ respectively. Then one writes down the
N-point function following the standard prescriptions. However, when we desire
to study the duality transformation 
properties of the scattering amplitudes for the states
arising out of compactifications, we have to evolve a strategy. I have proposed
a procedure to construct suitably modified vertex operators so that we
can implement duality transformations on them. This is one of the main
results of this investigation. The basic point to recognize is
that the backgrounds which arise after compactification, let us focus on
massless scalars for definiteness, get transformed
 under T-duality. Moreover, it is convenient
to combine these backgrounds to define the $M$-matrix as is well known.
In the weak field approximation, for we need to adopt a suitable scheme, we
have recalled how the backgrounds are expanded. Therefore, we should adopt
a similar prescription for the $M$-matrix in the weak field approximation.
To recall the case of 'metric' $G_{\alpha\beta}$, we express it as
$G_{\alpha\beta}=\delta_{\alpha\beta}+h_{\alpha\beta}$. Similarly, we adopt
a scheme where we expand, $M={\bf 1}+{\tilde M}$. 
We also noted that ${\tilde M}$ is not an
$O(d,d)$ matrix. Therefore, if we desire to apply T-duality transformation 
to generate a new amplitude starting from a given one, we have to follow
the steps outlined in this article. Thus we start from an initial 
configurations of backgrounds and construct the $M$-matrix, then adopt
the weak field approximation. In order to employ weak field approximation
for transformed backgrounds, we should transform the $M$-matrix to 
generate the new matrix, $M'$ and then adopt the weak field approximation.
We have dealt with the technicalities involved and laid down the procedures. 
Moreover, we presented
a simple example for $T^2$ compactification and showed that starting from
only the scalars coming from the compactification of the metric, we can
generate the scalars associated with the $B$-field by following the steps
proposed in this work. We also argued that following the arguments of
\cite{jmmass} we can implement T-duality transformation on vertex operators
associated with the massive scalars. We have found that the prescriptions of
\cite{klt} are quite efficient and elegant to construct T-duality invariant
vertex operators and implementation of duality transformation becomes quite
transparent. Moreover, we have studied a four point function where a gauge
boson scatters with a tachyon. The closed string gauge boson, we termed it
as gravi-photon, arises from compactification of the $\hat D$-dimensional
metric. We argued that under T-duality this scattering amplitude can be
related to an amplitude where the final state is an admixture of gravi-photon
and axi-photon the latter arises from compactification of the $B$-field.
I feel that the present investigation is another application of the T-duality
symmetry in the context of scattering of stringy states.
\\
Let us discuss a few possible applications of the present work. In recent
years, there have been interests in the study of massive excited states
in string theory \cite{heavy1,heavy2,heavy3,heavy4,heavy5,heavy6}.
 Highly excited massive states have exponential
degeneracy and if these states have high enough mass they can be identified
with the stringy black holes. Such states have attracted a lot of attention
from diverse perspectives. It is also well known that, in certain string
theoretic models where some extra compact dimensions 
are permitted to be large. There is the prospect
experimentally observing these stringy states in currently operational
accelerator like LHC \cite{larger1,larger2}. 
In my opinion some of these studies might lead to
interesting result if T-duality is incorporated as an  additional ingredient.
We have also proposed that vertex operators of massive stringy states
of NSR string can be expressed in T-duality invariant form. However, we
succeeded in the case of massive levels in the  NS-NS sector. The present
work can be extended to NS-NS sector of superstring. Some of these 
investigations are underway and these will be presented in future.

\bigskip
\noindent
{\bf Acknowledgments:} I have benefited from discussions with 
Professor Sunil Mukhi
and I thank him for bringing reference \cite{mp} to my attention.  
I would like to thank
Dr. B. R. Das for his extraordinary cares and hospitality
which made this investigation possible. This work is supported by the People of 
the Republic of India through a Raja Ramanna Fellowship of DAE.

\end{document}